\documentclass[aps,prl,twocolumn,superscriptaddress]{revtex4}
\usepackage{amsmath}
\usepackage{amssymb}
\usepackage{graphicx}
\usepackage{footnote}
\usepackage{subfigure}
\usepackage{sidecap}
\usepackage{verbatim} 
\usepackage{microtype}
\usepackage{hyperref}
\usepackage{float}
\usepackage{bm}
\usepackage{setspace}

\begin{document}

\title{Topology, Length Scales and Energetics of Surfactant Micelles}
\author{Subas Dhakal}
 
\affiliation{Department of Biomedical and Chemical Engineering, Syracuse University, Syracuse, NY 13244}
\author{Radhakrishna Sureshkumar}
\email{rsureshk@syr.edu }
\affiliation{Department of Biomedical and Chemical Engineering, Syracuse University, Syracuse, NY 13244}
\affiliation{Department of Physics, Syracuse University, Syracuse, New York 13244, USA}
\date{\today}

\begin{abstract}

We investigate the morphology and energetics of a self-associating model cationic surfactant in water using coarse-grained molecular dynamics simulations. We develop an algorithm to track micelle contours and quantify various microstructural features, such as contour length, persistence length, and mesh size.  We demonstrate that branched and multiconnected structures govern the anomalous dependence of zero-shear viscosity on salt concentration. We predict reliably the end-cap energy of micelles, for the first time, directly from the simulations.
\end{abstract}

\maketitle
In aqueous solutions, surfactant molecules spontaneously self-assemble into diverse geometrically complex and dynamically fluctuating morphologies. It has long been known that the emerging superstructures can range from spherical and elongated cylindrical to very long, flexible wormlike  micelles with or without branches \cite{Debye51,Lequeux96,CatesCandau90,Danino95,Appell92,Magid97,Sangwai11,Lin94,Cardiel13} and topologically rich knotted structures \cite{Lin94,Cardiel13}. The diversity in microstructure and rheological properties make micellar solutions beneficial to numerous applications \cite{Maitland00} as hydrofracking fluids in oil industry, turbulent friction drag reducing agents \cite{Lin94}, thickening agents in consumer products, drug carriers in targeted delivery \cite{Weitz09}, and templates to create functional nanofluids with tunable mechanical or optical properties \cite{Nettesheim08, Helgeson10b, Cong11}.

Since the early work of Debye \cite{Debye51}, the microstructural transitions in micellar solutions have been   investigated both theoretically \cite{Israelachvili76,Mackintosh90,Shaul82,CatesCandau90,Odijk91} and experimentally \cite{Debye51,Danino95,Shikata94,Khatory93}. It is now well-recognized that the molecular structure of co-surfactant or salt has a spectacular effect on the morphology. In particular, aromatic organic salts have stronger binding affinity to the micelles and induce enormous growth. Consequently, micelles become very long, flexible and entangle even at relatively low surfactant concentration $c_D$. Dilute solutions with spherical or short cylindrical structures exhibit Newtonian fluid rheology \cite{Shikata94, Khatory93, Vasudevan08}. In contrast, solutions above the overlap concentration $\phi^{*}$ consist of very long-thread like structures with contour lengths that span from a few 10s of nm to several $\mu $m, and show viscoelastic behavior reminiscent of flexible polymer solutions \cite{CatesCandau90,Shikata94,Khatory93}. However, unlike polymers, WLMs can merge and undergo reversible breaking at time scales that are detectable both in scattering experiments and simulations. Under non-equilibrium conditions, such as, under shear flow, the structure, dynamics, and the resultant rheological properties could change dramatically: two notable examples are the shear induced structure  (SIS) formation, and  shear banding  \cite{Khatory93,Shikata94,Vasudevan08,Keller98, Gonzalez04,Yesilata06, Nicolas12, Fielding04, Miller07}. Due to such dynamical complexities, a quantitative description of the microstructure of micellar fluids is   incomplete. In this paper, we present a comprehensive simulation study of self-assembly, emerging structures, length scales, and the energetics of a model surfactant solution with explicit solvent, electrostatic and hydrodynamic interactions.
\begin{figure}
\includegraphics[width=2.75in]{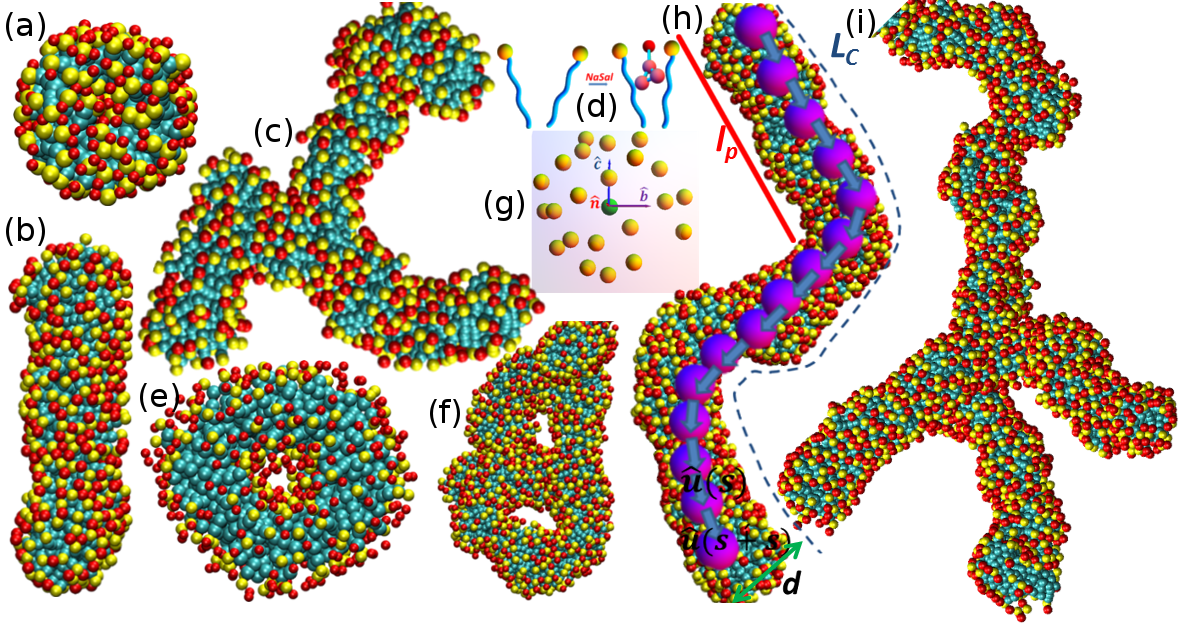}
\caption{\label{figure1}  (Color online) Different shapes of micelles observed in MD simulations. Color scheme: red (Sal$^-$), yellow (hydrophilic part of the surfactant), cyan (hydrocarbon tail). (a) Spherical. (b) Cylindrical. (c) \textbf{Y}-shaped micelle. (d)  Cartoon showing the effect of salt on relative orientation of two surfactant molecules.  (e-f) Micelles with handles, the number of handles depends on the physicochemical properties of the solution. (g) Schematic of Monge patch construction showing the charged head group of surfactants along with the approximated surface. (h) Wormlike micelle with $L \approx 50$ nm. (i) \textbf{X}-shaped micelle.}
\end{figure}

Despite decades of research aimed at understanding the structure and dynamics of micellar solutions, there are a few fundamental questions that still remain unanswered. First, although the scaling of average micelle length $\langle L \rangle$ for neutral or highly electrostatically screened micelles is clear \cite{CatesCandau90, Mackintosh90, Odijk91}, how $\langle L \rangle $ changes with $c_D$ and $c_s$  is not well understood at intermediate concentrations. Much remains unknown about the dependence of micelle length distribution on   $R = c_s/c_D$. In fact, contour length measurements of WLMs in the entangled regime by Cryo-TEM as well as light and neutron scattering experiments  have not yielded a conclusive topological picture \cite{Lequeux96}.  Second, the observed  non-monotonic dependence with multiple maxima  of zero-shear viscosity $\eta_0$ on $c_s$ \cite{ Oelschlaeger10, Hoffmann94} cannot be rationalized within the framework of the existing theories. Third, the end-cap  energy $E_c$ defined as excess energy of surfactants  in the hemispherical region ($E_s$) compared to the energy of those on the cylindrical body ($E_{cyl}$) of a micelle, is generally inferred indirectly from rheological measurements \cite{Helgeson10b,Larson14a,Nettesheim08}. To help address these fundamental questions, we simulate aqueous complexes of cetyltrimethylammonium chloride (CTAC) surfactants and strongly binding aromatic counterions--Sodium Salicylate (NaSal). From these simulations, we enumerate micelle morphologies, outline a phase diagram, quantify the end-cap energy and uncover the relevant length scales that help shed light on the anomalous viscosity variations with respect to $R$. Simulations utilize the coarse-grained MARTINI force field \cite{Marrink07, Sangwai11} and are performed using the LAMMPS software \cite{Plimpton95} in a constant NVT ensemble: see Supplemental Material \cite{supp} for details of the simulations and a micelle contour tracking algorithm developed to quantify the length scales, $\langle L \rangle$,  $l_p$, and $\xi$.  Topologically rich structures that emerge at different $R$ are shown in Fig. \ref{figure1}. 
\begin{figure}
\includegraphics[width=2.75in]{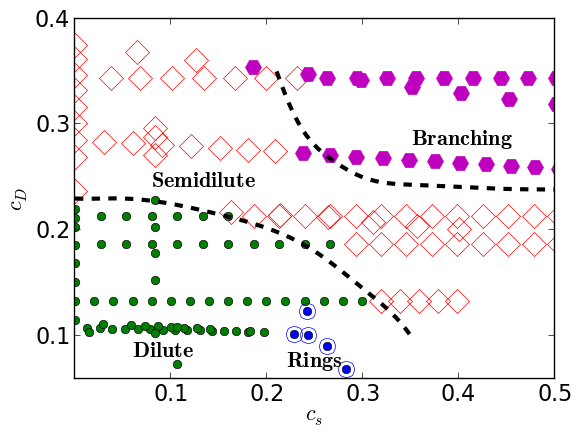}
\caption{ \label{figure2}(Color online) Phase diagram in terms of $c_s$ and $c_D$. The dotted lines are guides to the eye.}
\end{figure}

\emph{Phase diagram.} Figure \ref{figure2} outlines the phase diagram as functions of $c_D$ and $c_s$. We determine the boundary separating the dilute and semidilute regimes from the overlap volume of the micelles. Assuming each micelles as a rigid cylinder, we compute the volume fraction $\phi\equiv \sum_i^{N_m}L_i^3 / V$, where $L_i$, $N_m$ and $V$ are the micelle length, number of micelles, and the volume of the simulation box respectively.  A solution with $\phi<1$ is considered dilute, while a solution above the overlap concentration $\phi\geq\phi^{*}\approx 1$ consists of wormlike chains and  is semidilute. On further increasing the concentration, branched or multiconnected structures form. To determine the boundary between the semidilute and branched regimes in the phase diagram, we calculate the number of nodes $N_n=\sum_i^{N_m} N_n^i$ in the solution. Solutions with $30\%$ of the micelles having branches on average are considered as the branched phase. In a limited range of intermediate concentrations, we find micelles with handles [see Fig. \ref{figure1}(e-f)] and is shown by rings in Fig. \ref{figure2}. MD trajectory \cite{supp} clearly demonstrates that toroidal micelles form via end-cap attachment of a flexible cylindrical micelle.
\begin{figure}
\includegraphics[width=2.75in]{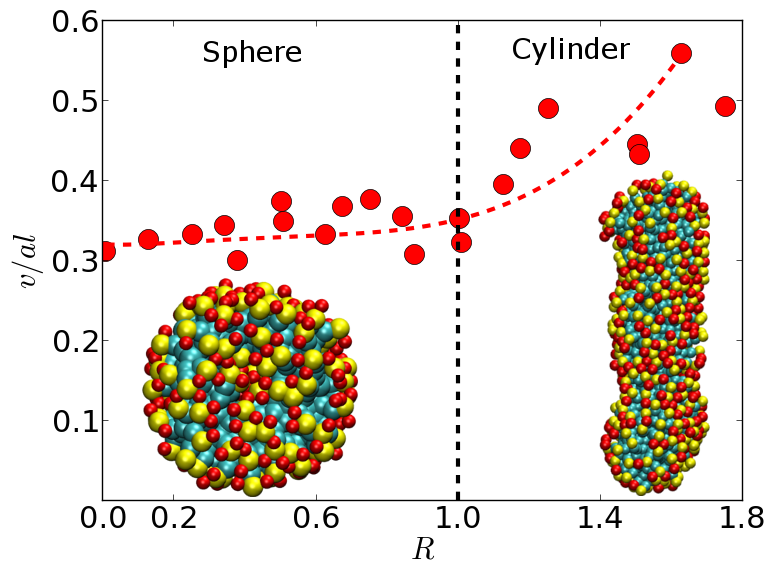}
\caption{\label{figure3}  (Color online) $v/{al}$ as functions of $R$ for $c_D$=0.10 M. Corresponding micelle shapes are also shown. The dotted line is given as a guide to the eye.}
\end{figure}

\emph{Shape transitions.}  It has been shown in earlier simulation studies \cite{Sangwai11} that  micelle shape crucially depends upon the salt concentration.  The mechanisms of this shape transformation can also be quantified by analyzing the free energy of an interfacial surfactant film when the film thickness is small compared to the neighboring layers. Mathematically, a surface embedded in $\mathbb R^3$ can be fully characterized by the mean $H=\frac{1}{2}\left(\frac{1}{R_1}+\frac{1}{R_2}\right)$, and the Gaussian $K=\frac{1}{R_1 R_2}$ curvatures. In the small curvature limit, the free energy of the film is given by
\begin{equation}\label{helfrichenergy}
F = 2 \kappa (H-H_0)^2+\bar{\kappa} K
\end{equation}
where $\kappa$ and $\bar{\kappa}$ are the bending moduli and $H_0$ is the spontaneous curvature of the film. For an amphiphile with surface area $a$ and liquid hydrocarbon volume $v$, the packing parameter that minimizes the free energy in Eq. \ref{helfrichenergy} is given by \cite{Israelachvili76}
\begin{equation}
\frac{v}{a l}= 1- \frac{l}{2} \left(\frac{1}{R_1}+\frac{1}{R_2}\right) +\frac{l^2}{3 R_1 R_2},
\end{equation}
where $l\approx 2.0 nm$  is the length of the hydrocarbon tail. Focusing on the micelles shapes, we numerically calculate $v/{a l}$ by mapping a small section of the interface to a Monge patch as illustrated in Fig. \ref{figure1}(g) which is fully described in SI \cite{supp}. It is interesting to note that the normals at vertices are unambiguously defined from the orientations $\textbf{n}$ of the amphiphilic molecules even at the  umblic  points. The average value of $v/{al}$  from these calculations  over a range of $c_s$ is shown in Fig. \ref{figure3}.  Without added salt or at lower salt concentrations, $v/al\approx 0.33$  and the thermodynamically most favorable shape is a sphere. Upon increasing $c_s$, Sal$^-$ ions interdigitate into the micelles surface [see Fig. \ref{figure1}(a-f)], thereby effectively screening the electrostatic interaction between the charged head groups [see Fig. \ref{figure1}(c)].  This effect gradually increases with increasing $c_s$ as shown by the decreasing effective surface area per surfactant in Fig. \ref{figure3}. In other words, the splay configuration costs more energy $\sim \left(\bm{\bigtriangledown}\cdot \textbf{n}\right)^2$ as compared to the uniform orientation of the molecules. This results in a more tightly packed  structure leading to micelle growth along the major axis, and cylindrical micelles are energetically favorable. 
\begin{figure}
\includegraphics[width=2.75in]{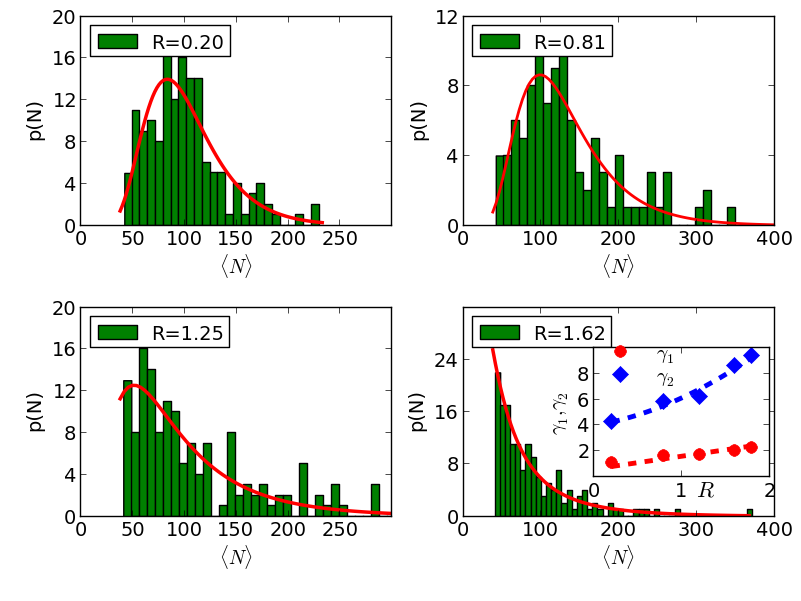}
\caption{\label{figure4} (Color online) Distributions of aggregation number $\langle N \rangle$ for different $R$. Inset shows the skewness ($\gamma_1$), and krutosis ($\gamma_2$) of the distributions with $R$. The lines are the fits to the distributions.}
\end{figure}

\emph{Micelle length distribution p(N).} To provide insights into micelle length distributions, we simulate systems with various $R$ at a fixed $c_D=0.16 M$. These simulations consist of $\approx$ one million atoms in a cubic box of dimensions $54 nm \times 54 nm \times 54 nm$. We conducted these simulations long enough ($t>700 ns$) to capture the breakage and recombination dynamics of the micelles \cite{supp}. Figure \ref{figure4} shows the distribution of aggregation number $N$ for different $R$. When $R>1$, the distribution $p(N)$ is an exponentially decaying function  as shown in Fig.\ref{figure4}(c-d). However, $p(N)$ is log-normal rather than exponential for $R < 1$ as evidenced from Fig.\ref{figure4}(a-b). For $R<1$, the Sal$^{-}$ ions condense non-uniformly over the micelles. It is possible, therefore, that electrostatic interactions in some micelles is only partially screened as compared to certain others. These results can be compared with the existing theories of Ref. \cite{CatesCandau90, Mackintosh90}, which argued that the interplay between entropy and the end-cap energy gives a broad, exponential distribution of lengths $p(N)\propto \exp\left(-N/\langle N\rangle\right)$. Our results clearly show that this is not the case for charged micelles, at least when $R<1$. Therefore, a more general distribution is suggested as:

\begin{equation}
p(N)\propto \frac{1}{\sqrt{2 \pi} \sigma  N} \exp \left(-\frac{\left[\ln(N) - \langle N \rangle \right]^2}{2 \sigma^2} \right).
\end{equation}

 The length distribution has a longer tail for a larger $R$. To quantify these effects, we show the $3^{rd}$ and $4^{th}$ moments ($\gamma_1$, and $\gamma_2$ respectively) in the inset plots, and both increase with $R$. 

\begin{figure}
(a)~\subfigure{\includegraphics[width=2.75in]{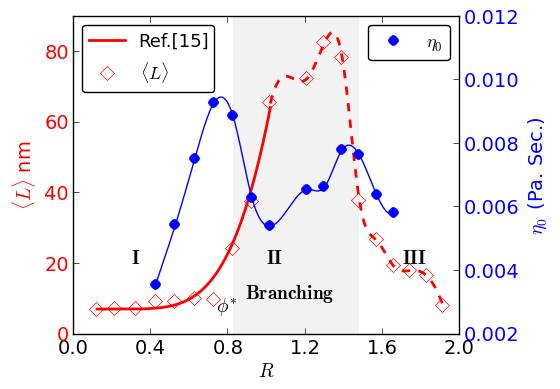}}
(b)~\subfigure{\includegraphics[width=2.75in]{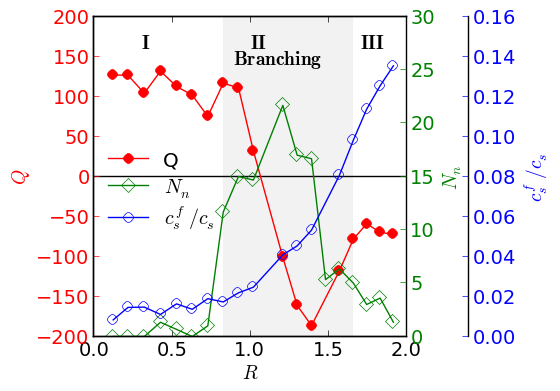}}
\caption{\label{figure5}  (Color online) (a) $\langle L \rangle$ and $\eta_0$ , and (b) the effective micelle charge Q, number of nodes $N_n$, and molar fraction of unbound counterions $c_s^f/c_s$ as functions of $R$ for $c_D$=0.32 M system in a simulation box of  size $30nm \times 30nm \times 30nm$. All lines except the thick solid red in (a) are guides to the eye.}
\end{figure}

\emph{Contour length $\langle L \rangle$ and zero-shear viscosity $\eta_0$.}  To understand the morphological changes underlying the non-monotonic variation of $\eta_0$ with $c_s$, we simulate systems with $c_D=0.32$M for $0\leq R \leq 2$. Figure \ref{figure5}(a) plots $\langle L \rangle$ and $\eta_0$ vs. $R$. Broadly, we can classify these variations into three regimes. Region I is the growth regime, which is further divided into two sub-regimes: those below and above $\phi^{*}$. For $\phi<\phi^{*}$, there is a gradual rise of $\langle L \rangle$ with increasing $R$. However, $\langle L \rangle$ increases sharply for $\phi \geq \phi^*$ in the semidilute regime. We interpret these results using an existing theory of micelle growth by MacKintosh et al. \cite{Mackintosh90} for charged micelles which predicts that $\langle L \rangle \propto 2 \phi^{1/2} \exp(E_c^\prime/{k_B T})$, where $E_c^\prime = \frac{1}{2}\left(E_c-\frac{l_B {v^*}^2 r}{\sqrt \phi}\right)$ is the renormalized end-cap energy, and $r$, $l_B$ and $v^*$ are the radius of micelles $\approx 2.35 nm$, Bjerrum length and the effective charge per unit length of micelles respectively. To compare our simulations with Ref. \cite{Mackintosh90}, we calculate the Bjerrum length $l_B\approx e^2/{\epsilon k_B T}\approx 0.88 nm$ from the radial distribution function of the charge \cite{supp}. Therefore, the effective charge per unit length of micelles $v^*\approx1/{2~l_B}=0.56 nm^{-1}$ \cite{Odijk91}. Using these estimates, we compare the predictions of the theory (thick solid line) with simulations (diamonds) in Fig. \ref{figure5}(a), and a very good agreement is found. In region II, as the branched or interconnected structures form, the contour length increases slightly  and attains a maximum when a fully saturated network forms. Finally, upon further increasing $c_s$ (region III), the increased entropy of free Sal$^{-}$ ions in the solution favors more end-caps and $\langle L \rangle$ decreases again. 
\begin{figure}
\includegraphics[width=2.75in]{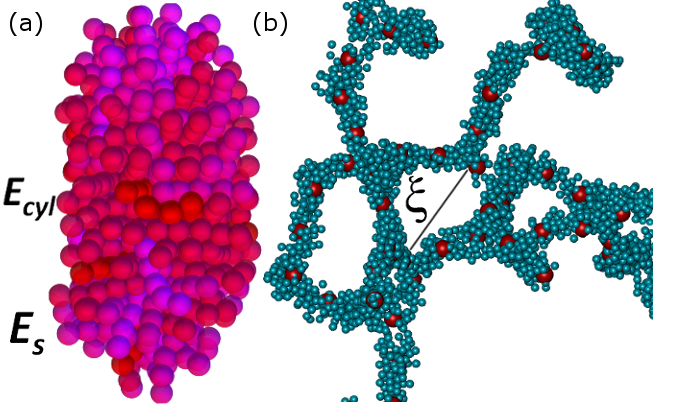}
\caption{ \label{figure6}(Color online) (a) Color map of surfactants in a micelle based upon chemical potential. (b) 2D view of a multiconnected network, while a 3D view is shown in SI \cite{supp}. }
\end{figure}

A key contribution of this study is uncovering the mechanisms that cause the anomalous variations of $\eta_0$ with respect to $R$ shown in Fig. \ref{figure5}(a). We extract $\eta_0$ from reverse non-equilibrium molecular dynamics (RNEMD) simulations \cite{Plathe99,supp}. Interestingly, $\eta_0$ exhibits two maxima as observed in experiments for similar systems \cite{Hoffmann94, Oelschlaeger10}. To further understand the underlying structural changes, we show the average micelle charge Q, fraction of unbound counterions $c_s^f/c_s$, and the node density $N_n$ in Fig. \ref{figure5}(b). We can correlate these data and micelle configurations with the viscosity changes. Initial increase in viscosity and $\langle L \rangle$ in region I can be attributed to the transition from spherical or short cylindrical to wormlike micelles. Beyond this maximum, as shown in Fig. \ref{figure5}(b), because of the condensed counterions the effective micelle charge becomes negative, promoting favorable inter-micelle interactions. Consequently, as shown in Fig. \ref{figure5}(a), viscosity  decreases for $0.8<R<1.0$ due to a transition from linear to branched micelles. Closer inspection of the microstructure reveals that some of the micelles are simply branched ($\textbf{Y}$ junctions) or have multiple branches while some are even cross-linked ($\textbf{X}$ junctions). This fact is illustrated by an increase of $N_n$ shown in Fig. \ref{figure5}(b). Such microstructures have previously been reported for several surfactant systems\cite{Danino95, Appell92,Porte86}. In fact, these structures are less viscous than entangled networks \cite{Appell92} of WLMs, and offers a faster mechanism for stress relaxation by sliding the  cross-links along the contour. Beyond the viscosity minimum at $R\approx 1.0$, micelle branches start to merge forming loops, resulting in a multiconnected network as shown in Fig. \ref{figure6}(b) of approximate mesh size $\xi\approx 21 nm$. The viscosity increases upon network formation, and passes through a second maximum at $R\approx 1.5$ when the network is fully saturated as signified by a maximum in $N_n$ in Fig. \ref{figure5}(b). The increase in viscosity from the branched state to the interconnected one is likely due to the increase rigidity of the network. Finally, in region III, the fraction of unbound counterions increases nonlinearly \cite{Shikata89} with increasing $R$, and the increased electrostatic attraction between the free counterions and the bound surfactants results in a gradual disintegration of the network as evidenced by a decrease in $N_n$. Consequently, micelles with branches or linear micelles are formed again. Hence, our simulations show that  the following sequence of morphological transitions manifests as the double maxima observed for $\eta_0$ vs. $c_s$ in Refs. \cite{Hoffmann94, Oelschlaeger10}: spherical $\rightarrow$wormlike ($\eta_0\uparrow$)$\rightarrow$branched ($\eta_0\downarrow$)$\rightarrow$ multiconnected ($\eta_0\uparrow$)$\rightarrow$branched/wormlike ($\eta_0\downarrow$).

\begin{figure}
(a)~\subfigure{\includegraphics[width=2.75in]{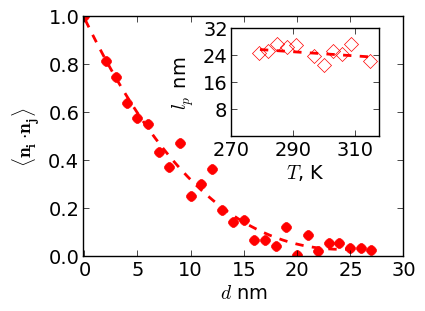}}
(b)~\subfigure{\includegraphics[width=2.75in]{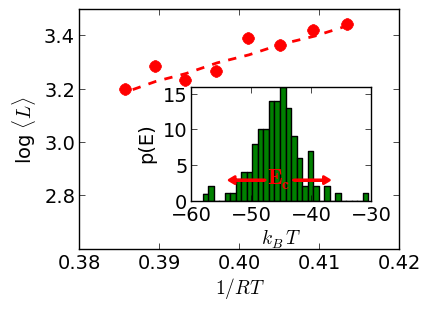}}
\caption{\label{figure7}(Color online) (a) Orientational correlation along the contour of micelles at $300 K$. Inset plots $l_p$ as functions of $T$. (b) $\langle L \rangle$ as a function of $T$. Inset shows the distribution of chemical potential of surfactants in a cylindrical micelle. These simulations correspond to $c_D$=$c_s$=0.20 M, and a simulation box of size $54nm \times 54nm \times 54nm$.} 
\end{figure}

\emph{Persistence length $l_p$.} The global flexibility $ x=\frac{l_p}{ \langle L \rangle} $  determines the viscoelastic properties of WLM solutions, where $l_p$ is the distance over which  orientational correlations are lost. Flexible behavior is observed for $x\ll 1$, whereas for $x\approx 1$ the micelles behave as rigid rods exhibiting I-N transition under shear flow \cite{Berret92,Gonzalez04}. To a first approximation, 1D bending modulus is defined as  $\kappa\sim l_p k_B T$. From the constructed contour path of the WLM chain, as demonstrated in Fig. \ref{figure1}(h), we define the orientation vector along the contour. Figure \ref{figure7}(a) shows the tangent to tangent correlations $\langle\textbf{n}(s) \cdot \textbf{n}(s+s\prime)\rangle =\exp\left(-s\prime/l_p\right)$, and the calculated $l_p$ at different $T$.  It is remarkable that the predicted value $l_p \approx 20 nm$, and $\kappa \approx 200 k_B T\AA$ are both consistent with experiments \cite{Helgeson10b,Oelschlaeger10,Larson14a}. Additionally, the inset shows that $l_p$ is a constant over the temperature range of $280-315 K$.

\emph{End-cap Energy $E_c$.} We present calculations  of $E_c$ by two methods. In the first method, we extract it using the theory of Mackintosh et al. \cite{Mackintosh90}.  Fig.\ref{figure7}(b) plots $\langle L \rangle$ vs. $1/{RT}$ and confirms the Arrhenius dependence as predicted in Refs. \cite{Mackintosh90, Helgeson10b}. From the slope of this plot we estimate an $E_c$ of $\approx 10 k_B T$, consistent with experiments \cite{Helgeson10b,Shikata94, Nettesheim08}. Second, we evaluate $E_c$ \emph{directly} by adding the pairwise interactions in simulations at $T=300 K$.  A color map of  the chemical potential of the surfactants in a typical cylindrical micelle is shown in Figure \ref{figure6}(a).  Inset shows the distribution of interaction energies of surfactants within a micelle, whose variance provides a direct measure of  $E_c \approx 12 k_B T$, which is in good agreement with the estimated value from the Arrhenius plot.

In summary, we have presented a quantitative analysis of the topology, length scales, and energetics of cationic surfactants micelle solution using extensive MD simulations. We have correlated the growth and branching of micelles with their viscosity, and proposed an underlying mechanism for the observed anomalies in viscosity upon increasing the salt concentration. Given that the model captures experimentally observed trends, we believe that framework presented here can offer an ideal platform to probe non-equilibrium phenomena in micellar fluids.
\acknowledgments
This work used the computational resources provided  by Science and Engineering Discovery Environment (XSEDE), which is supported by National Science Foundation grant number OCI-1053575. The authors acknowledged financial support by National Science Foundation under Grant No. 1049489 and 1049454.

\end{document}